\pdfoutput=1
\documentclass[reprint,
superscriptaddress,
 amsmath,amssymb,
 aps,
prb,
floatfix,
]{revtex4-2} 

\usepackage{graphicx}
\usepackage{dcolumn}
\usepackage{bm}
\usepackage{color}
\usepackage{ulem}

\begin{document}


\title{Auxiliary-field quantum Monte Carlo method with seniority-zero trial wave function}

\newcommand{\affilA}{%
Center for Quantum Information and Quantum Biology,
The University of Osaka, 1-2 Machikaneyama, Toyonaka, Osaka 560-0043, Japan
}%
\newcommand{\affilB}{%
Graduate School of Engineering Science, Osaka University, 1-3 Machikaneyama, Toyonaka, Osaka 560-8531, Japan
}%
\newcommand{\affilC}{Technology Strategy Center, TOPPAN Digital Inc., 1-3-3 Suido, Bunkyo-ku, Tokyo 112-8531, Japan}
\newcommand{\affilD}{Technical Research Institute, TOPPAN Holdings Inc., 4-2-3 Takanodai-minami, Sugito-machi, Kita-katsushika-gun, Saitama 345-8508, Japan}

\author{Yuichiro~Yoshida}
\email{yoshida.yuichiro.qiqb@osaka-u.ac.jp}
\affiliation{\affilA}
\author{Luca~Erhart}
\affiliation{\affilA}
\author{Takuma~Murokoshi}
\affiliation{\affilA}
\author{Rika~Nakagawa}
\affiliation{\affilC}
\author{Chihiro~Mori}
\affiliation{\affilC}
\author{Hanae~Tagami}
\affiliation{\affilD}
\author{Wataru~Mizukami}
\email{mizukami.wataru.qiqb@osaka-u.ac.jp} 
\affiliation{\affilA}%
\affiliation{\affilB}%

\date{\today}

\begin{abstract}
We present an approach that uses the doubly occupied configuration interaction (DOCI) wave function as the trial wave function in phaseless auxiliary-field quantum Monte Carlo (ph-AFQMC). DOCI is a seniority-zero method focused on electron pairs.
Although DOCI considers much fewer electron configurations than the complete active space (CAS) configuration interaction method, it efficiently captures the static correlation, while the consequent ph-AFQMC recovers the dynamical correlation across all orbitals. 
We also explore an orbital-optimized version (OO-DOCI) to further improve accuracy.
We test this approach on several chemical systems, including single O--H bond breaking in water and polymer additives. In these cases, OO-DOCI-AFQMC closely matches CAS-based ph-AFQMC and even outperforms coupled-cluster singles, doubles, and perturbative triples.
However, for strongly correlated systems, such as the carbon dimer and multi-bond dissociation in hydrogen systems and water, the method's accuracy drops. This suggests that seniority-zero space models may be insufficient as trial wave functions in ph-AFQMC for strongly correlated systems, suggesting the need for trial wave functions in an extended space.
Despite such a limitation, our study demonstrates that DOCI- and OO-DOCI-based ph-AFQMC can reduce the steep cost of CAS approaches, offering a path to accurate multi-reference calculations for larger, more complex systems.
\end{abstract}

\maketitle



\section{Introduction \label{sec:intro}}

Quantum chemical calculations have made significant advances in recent decades and have become standard practice in both computational chemistry and broader chemical research~\cite{Burke2012perspective, Jones2015density, Chanussot2021open, barroso_omat24, Merchant2023scaling}.
Single-reference (SR) methods, such as coupled-cluster singles, doubles, and perturbative triples (CCSD(T)), offer highly accurate descriptions of dynamical electron correlation for small to medium molecular systems~\cite{Smith2019approaching, Wilkins2019accurate}.
These methods are particularly effective in scenarios where the quasi-degeneracy of orbitals is not a critical factor~\cite{Lyakh2012multireference}. 
Even for large systems, such as biomolecules, high-precision calculations are now feasible due to the success of accelerated computational methods development~\cite{Riplinger2013efficient,Riplinger2013natural,Riplinger2016sparse,Saitow2017new,Guo2018communication,Liakos2020comprehensive,Guo2020linear}.
However, SR methods often fall short when dealing with systems that exhibit substantial static electron correlation, necessitating the ongoing development of multi-reference (MR) methods.

Accurate treatment of both dynamical and static electron correlation is crucial for MR systems, such as those encountered in molecular bond dissociation and transition metal complexes. 
Among the various approaches available, complete active space second-order perturbation theory (CASPT2)~\cite{Andersson1992second} is widely used to address these challenges. However, lower-order perturbation theories have limitations in capturing dynamical electron correlation, highlighting the need for continued research efforts to enhance methods for accurately representing this aspect of modern quantum chemistry.

Recently, the phaseless auxiliary-field quantum Monte Carlo (ph-AFQMC) method~\cite{Zhang2003quantum} has garnered significant attention for its ability to balance high accuracy with computational efficiency~\cite{Lee2022twenty}. This method, which operates through an imaginary-time evolution approach, has demonstrated successful applications to a wide range of systems, from molecules to solid state materials~\cite{Zhang2003quantum,Lee2022twenty,Taheridehkordi2023phaseless,Lee2019auxiliary,Neugebauer2023toward}.

A notable advantage of ph-AFQMC is its reliable description of dynamical electron correlation, coupled with high parallelization efficiency.
Various benchmark studies have validated its accuracy~\cite{Neugebauer2023toward,Sukurma2023benchmark,Wei2024scalable,Taheridehkordi2023phaseless,Lee2019auxiliary,Lee2022twenty}.
For example, Lee et al. showed that for datasets of SR molecular systems, ph-AFQMC is more accurate than coupled-cluster singles and doubles (CCSD) and less accurate than CCSD(T). Meanwhile, its computational cost scales from $\mathcal{O}(N^3)$ to $\mathcal{O}(N^4)$ per sample~\cite{Lee2022twenty}, where $N$ represents the number of orbitals.
This scaling offers a significant advantage over conventional approaches, where the computational costs for M{\o}ller-Plesset second-order perturbation theory (MP2), CCSD, and CCSD(T) scales as $\mathcal{O}(N^5)$, $\mathcal{O}(N^6)$, and $\mathcal{O}(N^7)$, respectively.
Moreover, graphics processing unit acceleration has recently been explored for ph-AFQMC~\cite{Huang2024GPU,Jiang2024improved}, further decreasing its computational time.
These features highlight the promise of ph-AFQMC as a method that balances accuracy and efficiency in treating dynamical electron correlation.

Another advantage of ph-AFQMC is that it naturally yields an MR wave function.
By using an MR trial wave function, ph-AFQMC can provide a balanced treatment of systems characterized by strong electron correlation~\cite{Zhang2003quantum}, making it particularly useful for problems beyond the scope of SR methods.
For example, by utilizing a complete active space self-consistent field (CASSCF) trial wave function, the ph-AFQMC method has showcased precise representations of multiple bond breaking in small systems, outperforming multi-reference perturbation theories and achieving outcomes on par with multi-reference configuration interaction~\cite{Lee2022twenty}.
Challenging applications to dinuclear transition metal complexes such as [Cu$_2$O$_2$]$^{2+}$ and [Fe$_2$S$_2$(SCH$_3$)]$^{2-}$ have also been investigated with MR trial wave functions~\cite{Mahajan2021taming,Huang2024GPU,Malone2023ipie}.
The capability of ph-AFQMC to handle both dynamical and static electron correlation renders it particularly appealing and deserving of further investigation.

However, a critical challenge arises from the exponential scaling in the preparation cost and size of the MR trial wave function in ph-AFQMC.
The typical method for generating an MR wave function, such as CASSCF, incurs exponential costs owing to the combinatorial explosion required to accommodate various electronic configurations.
Specifically, for a closed-shell system, the number of electronic configurations in the complete active space scales as $\binom{n}{k}^2$, where $n$ and $k$ represent the number of spatial orbitals and electrons per spin, respectively. 
This exponential growth significantly limits the feasibility of CASSCF and, by extension, the utilization of CASSCF trial wave functions within ph-AFQMC for larger systems.
In fact, several studies to perform AFQMC calculations using MR trial wave functions other than the complete active space (CAS) wave function have been reported in recent years~\cite{Mahajan2022selected,Mahajan2024beyond,Jiang2025unbiasing,Chang2024boosting,Huggins2022unbiasing,Huang2024evaluating,Sukurma2025self,Khinevich2025enhancing,Yoshida2025auxiliary,Danilov2025enhancing}, suggesting the importance of this line of research.

Doubly occupied configuration interaction (DOCI) and orbital-optimized DOCI (OO-DOCI) methods are utilized to construct MR wave functions by incorporating ground-state and pair-excited electronic configurations~\cite{Allen1962electron,Smith1965natural,Weinhold1967reduced,Veillard1967complete,Cook1975doubly,Fantucci1977direct,Couty1997generalized,Kollmar2003new}.
These so-called seniority-zero wave functions effectively capture static electron correlation~\cite{Bytautas2011seniority,Limacher2013new,Raemdonck2015polynomial}, particularly in systems with orbital degeneracy or bond-breaking scenarios.
Despite the exponential growth in preparation costs, the number of configurations is reduced to $\binom{n}{k}$ rather than $\binom{n}{k}^2$, significantly decreasing the computational prefactor and allowing for the treatment of larger active spaces.

In this study, we propose the use of DOCI and OO-DOCI wave functions as the trial wave function for ph-AFQMC to efficiently capture static electron correlation while addressing the high cost associated with MR trial wave function preparation.
These methods are referred to as DOCI-AFQMC and OO-DOCI-AFQMC, where the quantum chemical calculations used for the trial wave function are placed before AFQMC.
Their effectiveness is demonstrated in several systems, including hydrogen systems, H$_2$O, and the carbon dimer.
Furthermore, we showcase their applicability to industrially relevant molecular systems by investigating O--H bond dissociation in polymer additives~\cite{Lucarini2010free}.

The remainder of this paper is organized as follows.
Sec.~\ref{sec:methods} provides an overview of the ph-AFQMC and DOCI methods, introducing the formalism of (OO-)DOCI-AFQMC.
Section~\ref{sec:comput_details} outlines the computational procedures.
Numerical results are presented in Sec.~\ref{sec:results}, with concluding remarks on the validity and limitations of the proposed methods presented in Sec.~\ref{sec:conclusions}.

\section{Computational methods \label{sec:methods}}

This study presents the utilization of DOCI and OO-DOCI wave functions as trial wave functions for ph-AFQMC.
This section provides a brief review of both the ph-AFQMC and DOCI methods.

\subsection{Review of ph-AFQMC}

AFQMC is a method utilized to determine the electronic ground state $| \Psi_0 \rangle$ of a system through imaginary-time evolution:
\begin{align}
    | \Psi_0 \rangle = \lim_{\tau \to \infty} e^{-\tau \hat{H}} | \Phi_0 \rangle
\end{align}
where $|\Phi_0\rangle$ represents an initial wave function and $\tau$ is the imaginary time.
The operator $\hat{H}$ denotes the electronic structure Hamiltonian of the target system:
\begin{align}
    \hat{H} = \sum_{p,q} h_{pq} \hat{a}_p^\dagger \hat{a}_q + \frac{1}{2}\sum_{p,q,r,s} g_{prqs} \hat{a}_p^\dagger \hat{a}_q^\dagger \hat{a}_s \hat{a}_r,  \label{eq:H}
\end{align}
where $\hat{a}^\dagger_p$ and $\hat{a}_p$ represent creation and annihilation operators for the $p$th spin-orbital, and $h_{pq}$ and $g_{prqs}$ denote the one- and two-electron integrals, respectively.
The two-electron integral tensor is factorized in a way such as modified Cholesky decomposition~\cite{Beebe1977simplifications,Aquilante2010MOLCAS,Motta2018abinitio}
\begin{align}
    g_{prqs} \simeq \sum_{l=1}^{N_l} L_{pr}^l L_{qs}^l, \label{eq:Cholesky}
\end{align}
where $N_l$ is the number of Cholesky vectors.

By substituting Eq.~(\ref{eq:Cholesky}) into Eq.~(\ref{eq:H}), the Hamiltonian is expressed as
\begin{align}
    \hat{H} \simeq \hat{v}_0 - \frac{1}{2}\sum_{l=1}^{N_l} \hat{v}_l^2, \label{eq:H2}
\end{align}
where
\begin{align}
    \hat{v}_0 &= \sum_{p,q} \left( h_{pq} - \frac{1}{2} \sum_t g_{pttq} \right) \hat{a}_p^\dagger \hat{a}_q \\
    \hat{v}_l &= i \sum_{p,q} L_{pq}^l \hat{a}^\dagger_p \hat{a}_q.
\end{align}
In this way, the Hamiltonian is rewritten in a form that uses one-body operators effectively.

By applying the Trotter decomposition and Hubbard-Stratonovich transformation, the short-time propagator $e^{-\Delta\tau\hat{H}}$ is expressed as
\begin{align}
    \textcolor{black}{
    e^{-\Delta\tau\hat{H}} = \int d{\bm x}\,p({\bm x})\hat{B}({\bm x}, \Delta\tau) + \mathcal{O}(\Delta\tau^2),
    \label{eq:HS}
    }
\end{align}
\textcolor{black}{where $p({\bm x})$ is the normal distribution and ${\bm x}$ is an auxiliary-field vector. $\hat{B}({\bm x}, \Delta\tau)$ is an effective one-body propagator coupled to the auxiliary field ${\bm x}$ as}
\begin{align}
    \textcolor{black}{
    \hat{B}({\bm x}, \Delta\tau) := e^{-\frac{\Delta\tau}{2}\hat{v}_0} 
    e^{\sqrt{\Delta\tau}\sum_l x_l\hat{v}_l} 
    e^{-\frac{\Delta\tau}{2}\hat{v}_0}.
    }
\end{align}
\textcolor{black}{In this effective one-body propagator, the one-body operators $\hat{v}_l$, decomposed two-body term in the Hamiltonian, are coupled to the auxiliary-field.  
It can be considered that the free fermionic system evolves via the coupling to the auxiliary fields, and the many-body effect is recovered through a stochastic integral over the fields in Eq.~(\ref{eq:HS}).}


In ph-AFQMC, the ground-state wave function at imaginary-time $\tau$, $| \Psi (\tau) \rangle$, is represented as follows:
\begin{align}
    | \Psi (\tau) \rangle = \sum_{i=1}^{N_w} w_i \frac{|\Phi_i (\tau) \rangle}{\langle \Psi_{\rm T} | \Phi_i (\tau)\rangle}, \label{eq:psitau}
\end{align}
where $\{ |\Phi_i(\tau)\rangle \}$ represents a set of walkers. \textcolor{black}{Each walker is a single Slater-determinant in arbitrary single-particle basis and updated by applying the propagator as}
\begin{align}
    \textcolor{black}{| \Phi_i(\tau + \Delta\tau) \rangle = \hat{B}({\bm x}_i, \Delta\tau) | \Phi_i(\tau) \rangle.}
\end{align}

A trial wave function $|\Psi_{\rm T}\rangle$ is introduced to address the phase problem in the walker weights $\{ w_i \}_{i=1}^{N_w}$.
In brief, the weight of each walker is updated based on the overlap ratio $S_i(\tau)$:
\begin{align}
    w_i(\tau + \Delta\tau) \propto w_i(\tau)  \textcolor{black}{|S_i(\tau)| \max\{0,\,\cos(\Delta\theta_i)\}} ,
\end{align}
where $S_i(\tau)$ is defined as follows:
\begin{align}
   S_i(\tau) = \frac{\langle \Psi_{\rm T} | \Phi_i(\tau + \Delta\tau) \rangle}{\langle \Psi_{\rm T} | \Phi_i(\tau) \rangle}.
\end{align}
To mitigate the phase problem, the weight is adjusted by multiplying by $\max\{0,\, \cos(\Delta\theta_i)\}$, where $\Delta\theta_i = \arg[S_i(\tau)]$. 
This adjustment effectively eliminates negative contributions that lead to numerical instabilities, as $\cos(\Delta\theta_i)$ corresponds to the real component of the overlap ratio.

By constraining the phase in this manner, ph-AFQMC prevents the uncontrolled growth of the phase problem, albeit at the cost of introducing a reliance on the trial wave function. Consequently, the accuracy of ph-AFQMC calculations is significantly influenced by the choice of $|\Psi_{\rm T}\rangle$.

\subsection{Review of DOCI}

DOCI is a configuration interaction (CI) method that constructs an MR wave function using the ground-state configuration and its pair-excited configurations~~\cite{Weinhold1967reduced,Veillard1967complete,Cook1975doubly,Couty1997generalized,Kollmar2003new}. The DOCI wave function is expressed as follows:
\begin{align}
    |\Psi_{\rm DOCI}\rangle= \sum_{i=1}^{N_{{\rm DO}}} c_i \hat{s}_i |\rangle, 
\end{align}
where $|\rangle$ denotes the vacuum state, $N_{\rm DO}$ represents the total number of doubly occupied (DO) electronic configurations, $c_i$ represents the expansion coefficient of the $i$th DO configuration, and $\hat{s}_i$ denotes the corresponding creation operator. Specifically, 
\begin{align}
    \hat{s}_i = \prod_{j=1}^{n} \left (\hat{a}^\dagger_{j\alpha} \hat{a}^\dagger_{j\beta}\right )^{f_{ij}},
\end{align}
where $\hat{a}^\dagger_{j\sigma}$ generates an electron in the $j$th spatial orbital with spin $\sigma$.
$n$ represents the number of spatial orbitals, and $f_{ij} \in \{0, 1\}$ indicates whether the $j$th orbital is doubly occupied in the $i$th configuration.
\textcolor{black}{When an active space is introduced, the above operators are given only for the defined set of active orbitals and the configuration interactions involving the external orbitals are not considered.}

The total number of DO electronic configurations is expressed as:
\begin{align}
    N_{{\rm DO}} = \binom{n}{k},
\end{align}
where $k$ represents the number of electrons of a given spin.
Recall that the number of configurations in a complete active space for a closed-shell system is expressed as $\binom{n}{k}^2$. Therefore, DOCI reduces the configuration space to the square root of that in the complete active space, allowing for the treatment of larger active spaces with fewer configurations.

\begin{figure}[ht]
    \centering
    \includegraphics[width=1.0\linewidth, bb=0 0 610 213]{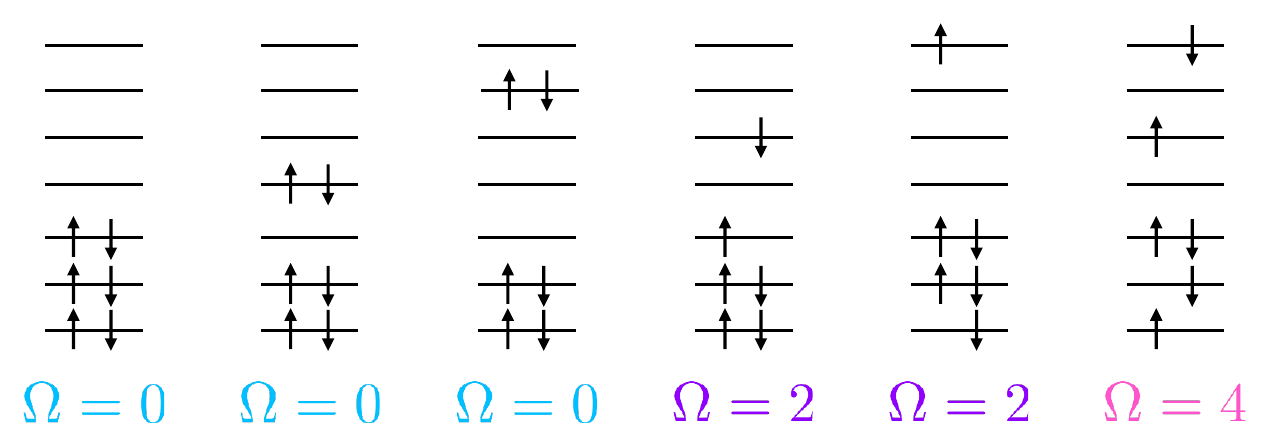}
    \caption{Example electronic configurations and their corresponding seniority numbers.}
    \label{fig:seniority}
\end{figure}

A DOCI wave function is constructed from electron configurations with a seniority number $\Omega = 0$.
The seniority number $\Omega$ is a measure of the number of unpaired electrons in a configuration~\cite{Bytautas2011seniority}.
An example of electronic configurations with different seniority numbers is shown in Figure~\ref{fig:seniority}.
In this example, the seniority-zero (DOCI) wave function can encompass the three configurations on the left while excluding those on the right (which have $\Omega \neq 0$).
The seniority-zero constraint effectively captures the primary electron excitations that dominate the ground-state character in many strongly correlated systems~\cite{Stein2014seniority,Henderson2014seniority,Henderson2015pair,Henderson2019geminal,Senjean2022reduced}.

\section{Computational details} \label{sec:comput_details}

This section provides an overview of the computational procedures and the software utilized.

PySCF v2.2.1~\cite{PySCF,PySCF2} was employed to perform self-consistent field (SCF), CCSD, CCSD(T), CASCI, and CASSCF calculations.
Additionally, DOCI and OO-DOCI calculations were performed using the DOCI module for PySCF v0.1~\cite{DOCI}.
The cc-pVDZ basis set was employed for all calculations unless otherwise specified.
Density functional theory calculations were conducted using Gaussian~16 Revision C~\cite{g16} for comparison.

Following the multi-configurational self-consistent field (MCSCF) calculations, ph-AFQMC calculations with the MR trial wave function were performed using ipie v0.6.2~\cite{Malone2023ipie}.

The total imaginary-time evolution $\tau_{\rm total}$ is defined as follows:
\begin{align}
    \tau_{\rm total} = N_b \times n_s \times \Delta \tau,
\end{align}
where $N_b$ and $n_s$ represent the number of blocks and time steps, respectively.
Each block serves as a unit for grouping computations to obtain energy expectation values and other observables.

Unless specified otherwise, the ph-AFQMC parameters were set as follows: time step $\Delta \tau = 0.005~E_h^{-1}$, number of steps $n_s = 50$, number of walkers $N_w = 400$, and number of blocks $N_b = 3000$.

\section{Results and discussion} \label{sec:results}

\subsection{Hydrogen chain and ring}

First, we present the results for a linear H$_4$ chain and hexagonal H$_6$ ring, both of which serve as simple models of strongly correlated electron systems~\cite{Sinitskiy2010strong,Kats2013Communication}.

\begin{figure}[ht]
    \centering
    \includegraphics[width=1.0\linewidth, bb=0 0 461 346]{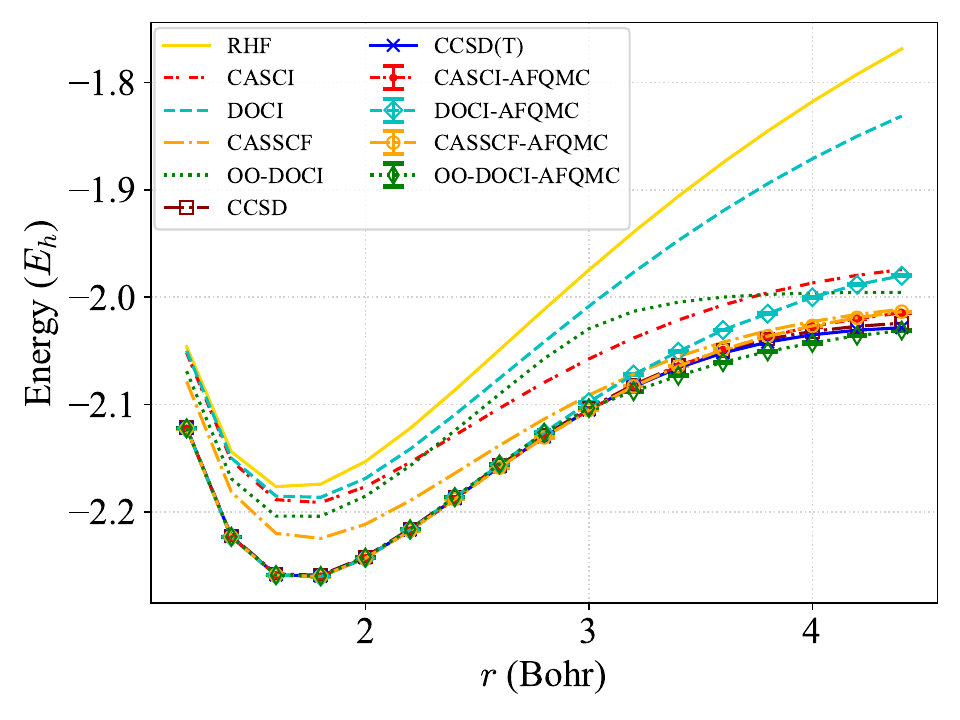}
    \caption{Potential energy curve of the linear H$_4$ chain using the cc-pVDZ basis-set. The active space is (4e, 4o), with the variable $r$ representing the distance between neighboring hydrogen atoms along the horizontal axis.}
    \label{fig:h4}
\end{figure}

The potential energy curve of the linear H$_4$ chain is shown in Figure~\ref{fig:h4}.
Overall, ph-AFQMC recovers the dynamical electron correlation that is absent in the bare MCSCF calculations. Particularly, OO-DOCI-AFQMC closely reproduced CASSCF-AFQMC and demonstrated good agreement with CCSD and CCSD(T).

By contrast, DOCI-AFQMC \textcolor{black}{under}estimates the correlation energy in the dissociation region ($r >$ 3.0~Bohr).
This discrepancy may be attributed to the limited flexibility of the DOCI trial wave function, as the original DOCI energy curve fails to accurately capture the behavior at large bond distances.

\begin{figure}[ht]
    \centering
    \includegraphics[width=1.0\linewidth, bb=0 0 461 346]{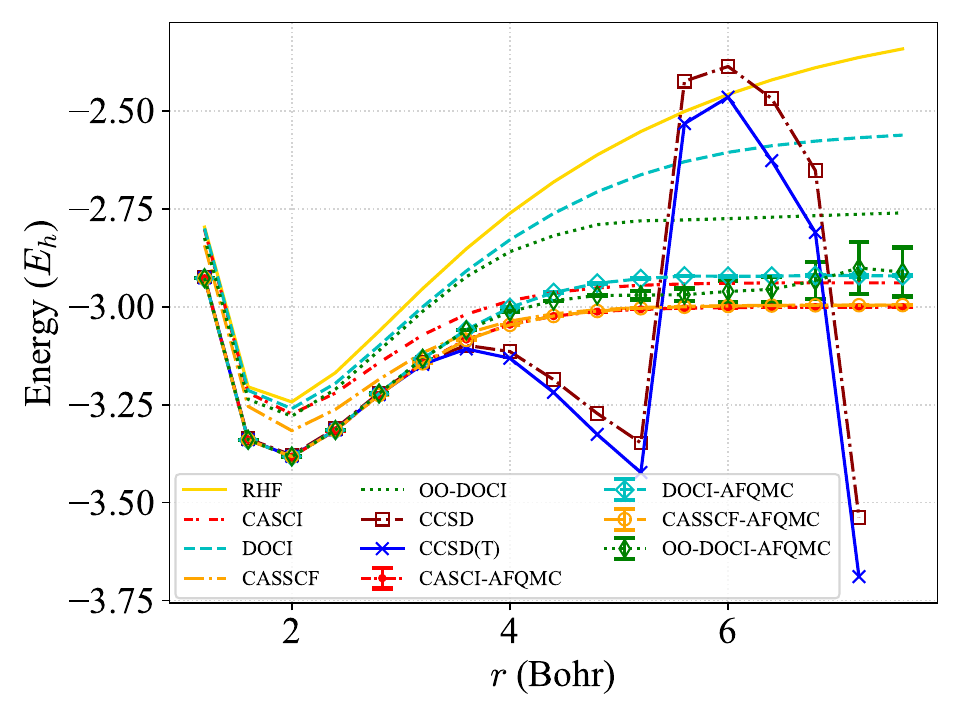}
    \caption{Potential energy curve of the hexagonal H$_6$ ring using the cc-pVDZ basis-set. The active space is (6e, 6o), with the variable $r$ representing the distance between neighboring hydrogen atoms on the horizontal axis.}
    \label{fig:h6}
\end{figure}

Next, we considered the H$_6$ ring, with the Cartesian coordinates of the $i$th hydrogen atom defined as follows:
\begin{align}
\bm{r}_i =
\begin{pmatrix}
x_i \\
y_i \\
z_i
\end{pmatrix} =
\begin{pmatrix}
r \cos\left(\frac{i \pi}{3}\right) \\
r \sin\left(\frac{i \pi}{3}\right) \\
0
\end{pmatrix},
\end{align}
where $i = 0, 1, \cdots, 5$. The resulting energy curves are shown in Figure~\ref{fig:h6}.
Similar to the H$_4$ case, ph-AFQMC recovers the missing dynamical electron correlation in the MCSCF calculations while capturing the strong static correlation essential for the ring dissociation.
Both DOCI-AFQMC and OO-DOCI-AFQMC, utilizing MR trial wave functions, provide qualitative descriptions of dissociation and compare reasonably with CASCI-AFQMC and CASSCF-AFQMC.
In contrast, the single-reference CCSD and CCSD(T) break down at large $r$~\cite{Kats2013Communication}.

\begin{figure}[ht]
    \centering
    \includegraphics[width=1.0\linewidth, bb=0 0 461 346]{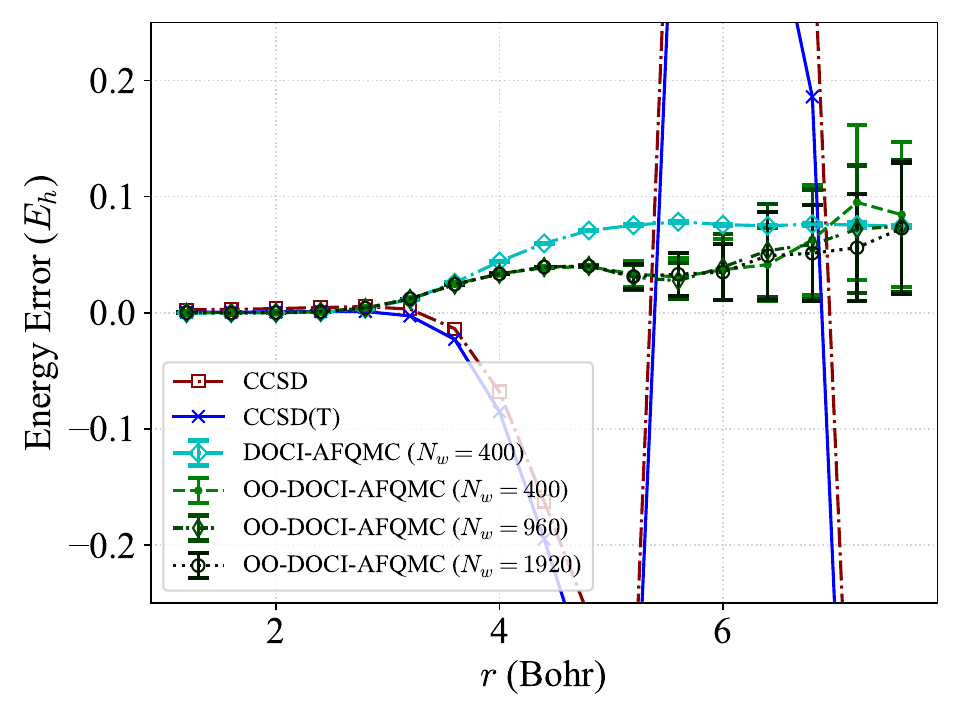}
    \caption{Energy errors relative to CASSCF-AFQMC for the H$_6$ ring using the cc-pVDZ basis set. The active space is (6e, 6o), with the variable $r$ representing the distance between neighboring hydrogen atoms along the horizontal axis.}
    \label{fig:h6_error}
\end{figure}

In the region of $r \geq 6$~Bohr, the standard error of OO-DOCI-AFQMC increased considerably. To investigate this \textcolor{black}{anomalous behavior}, we increased the number of walkers $N_w$; the results are shown in Figure~\ref{fig:h6_error}.
We plotted graphs of energy differences relative to CASSCF-AFQMC computed with $N_w=400$.
The standard error in these differences accounted for the propagated error from the reference CASSCF-AFQMC energies, which is sufficiently small (see Figure~\ref{fig:h6}).

Despite increasing $N_w$, the standard error of OO-DOCI-AFQMC remains largely unchanged and fails to address the issue at large values of $r$. 
Because the H$_6$ ring is driven by a strong static correlation, a seniority-zero wave function may not provide a fully quantitative description of the dissociation region.

\subsection{Two O--H bond dissociation in water}

Next, we examine the simultaneous stretching of the two O--H bonds in H$_2$O, a well-known model of double bond dissociation~\cite{Henderson2014seniority,Kinoshita2005coupled,Erhart2024coupled}.
An active space of (4e, 4o) was selected, comprising two $a_1$ and two $b_2$ orbitals while excluding the $b_1$ orbital dominated by the oxygen $p$-orbital perpendicular to the molecular plane.
The equilibrium geometry of H$_2$O, utilized as the starting point for varying the two O--H bond lengths, was optimized at the MP2=FULL/cc-pVDZ level~\cite{CCCBDB}.

\begin{figure}[ht]
    \centering
    \includegraphics[width=1.0\linewidth, bb=0 0 461 346]{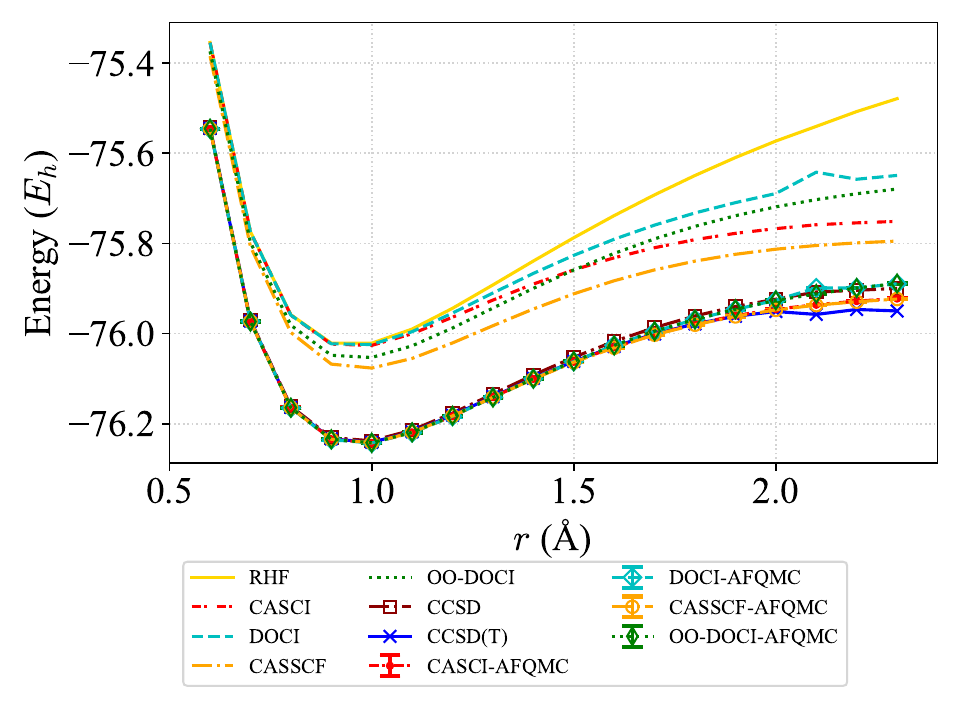}
    \caption{Potential energy curve of the simultaneous dissociation of two O--H bonds in H$_2$O utilizing the cc-pVDZ basis set. The active space is (4e, 4o), with the variable $r$ representing the dissociating O--H bond length along the horizontal axis.}
    \label{fig:h2o}
\end{figure}

The potential energy curve for the simultaneous O--H bond dissociation is shown in Figure~\ref{fig:h2o}.
All ph-AFQMC methods, whether utilizing CASSCF, CASCI, OO-DOCI, or DOCI trial wave functions, offer a reasonable description of the energy curve encompassing the dissociation region.
Neither ph-AFQMC, CCSD, nor CCSD(T) \textcolor{black}{showed an obvious break, but as $r$ increases, a change in energy begins to be seen}.

\begin{figure}[ht]
    \centering
    \includegraphics[width=1.0\linewidth, bb=0 0 461 346]{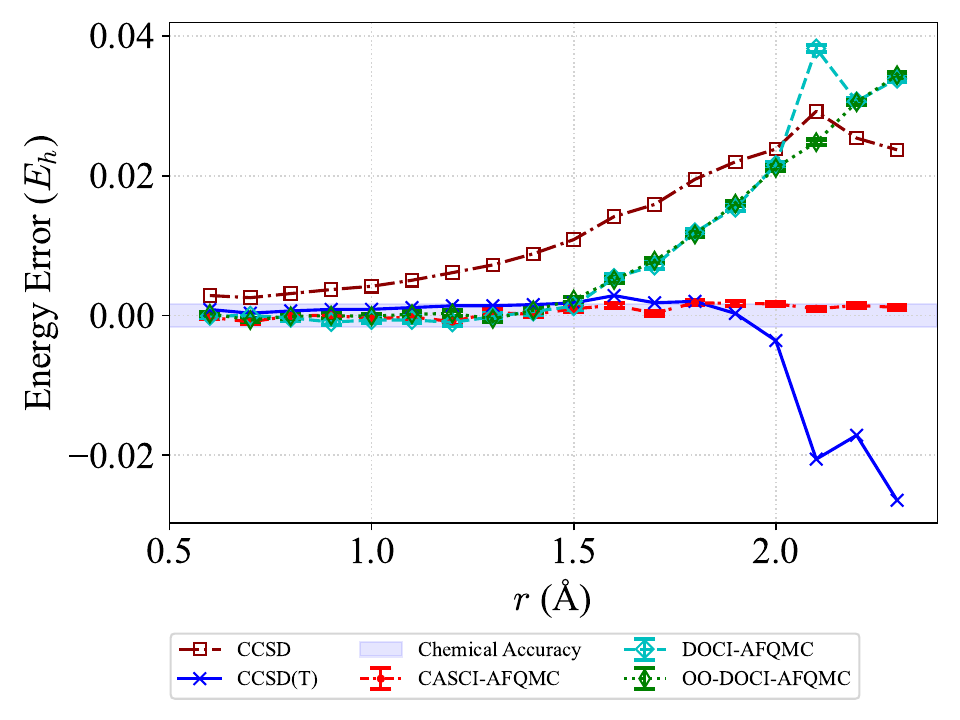}
    \caption{Energy deviations relative to CASSCF-AFQMC for two O--H bond dissociation in H$_2$O using the cc-pVDZ basis set. The active space is (4e, 4o), with the variable $r$ representing the dissociating O--H bond length along the horizontal axis.}
    \label{fig:h2o_error}
\end{figure}

\textcolor{black}{To emphasize the magnitude of the energy errors, Figure~\ref{fig:h2o_error} shows the deviations of each method from the CASSCF-AFQMC reference.}
DOCI-AFQMC and OO-DOCI-AFQMC maintain chemical accuracy in structures close to the equilibrium geometry ($r \lesssim 1.5$~\AA).
However, as $r$ surpassed 1.5~\AA, errors in both methods gradually increased, indicating that the seniority-zero trial wave functions may not be adequate for a precise description of this strongly correlated regime.

\textcolor{black}{CCSD and CCSD(T) show a similar trend to DOCI- and OO-DOCI-AFQMC in that the error begins to increase as $r$ increases. The difference between the complete active space and seniority-zero cases highlights the importance of the $\Omega > 0$ configurations in the trial wavefunction in the dissociation region.}

\subsection{Carbon dimer}

\begin{figure}[ht]
    \centering
    \includegraphics[width=1.0\linewidth, bb=0 0 720 576]{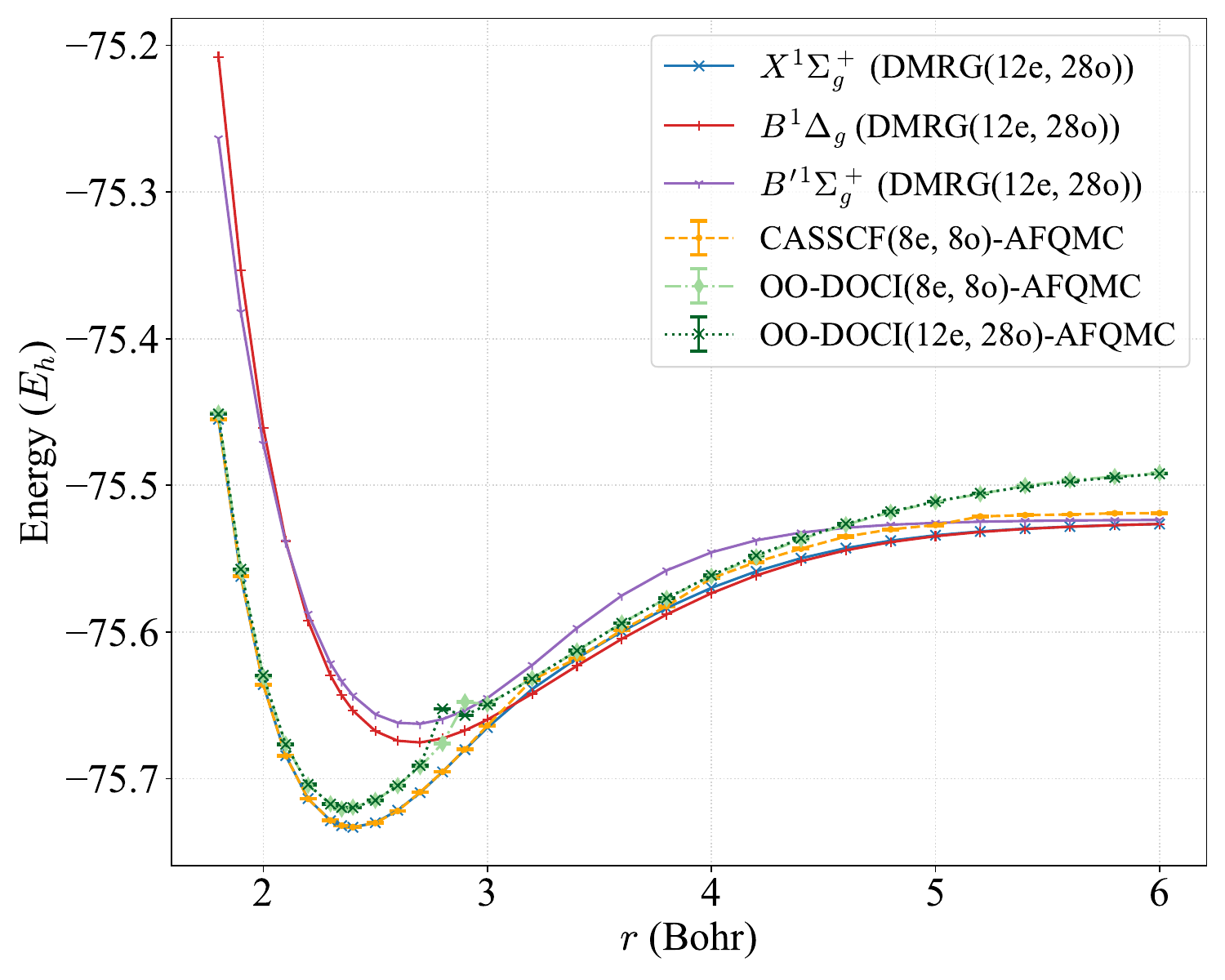}
    \caption{Potential energy curve of the carbon dimer using the cc-pVDZ basis set. The variable $r$ on the horizontal axis is the bond length.}
    \label{fig:c2}
\end{figure}

Subsequently, we delve into a more complex scenario: multiple bond dissociation in the carbon dimer. The potential energy curves obtained from ph-AFQMC using CASSCF(8e, 8o), OO-DOCI(8e, 8o), and OO-DOCI(12e, 28o) trial wave functions are shown in Figure~\ref{fig:c2}.
We \textcolor{black}{also show the} results from density matrix renormalization group (DMRG) calculations conducted by Wouters et al.~\cite{Wouters2014CheMPS2} for comparison.
Note that the active space (12e, 28o) includes all orbitals in the cc-pVDZ basis set.

While OO-DOCI-AFQMC captured the qualitative shape of the curve, it fell short in accuracy compared with CASSCF-AFQMC when referenced against the DMRG results.
Specifically, for $r < 3.0$~Bohr, CASSCF(8e, 8o)-AFQMC closely aligned with DMRG, whereas OO-DOCI(8e, 8o)-AFQMC demonstrated significant deviations.
Despite the larger active space accommodated by OO-DOCI(12e, 28o) compared with CASSCF, its energy profile closely mirrored that of OO-DOCI(8e, 8o)-AFQMC.
Achieving a more precise depiction of the carbon dimer necessitates the inclusion of configurations with $\Omega > 0$ in the MR trial wave function.

The DMRG outcomes revealed a crossing point between the $X{}^1\Sigma^+_g$ and $B{}^1\Delta_g$ states around a distance of approximately $r \sim 3.1$~Bohr, whereas an avoided crossing was observed between the $X{}^1\Sigma^+_g$ and $B'{}^1\Sigma^+_g$ states.
Both CASSCF-AFQMC and OO-DOCI-AFQMC demonstrated discontinuities near $r = 3.0$~Bohr, albeit at different points.
This region of crossing/avoided-crossing likely triggered the observed discontinuities, and \textcolor{black}{the similar discontinuity} has been reported in the computational investigations of frozen-pair \textcolor{black}{coupled cluster theory}~\cite{Leszczyk2022assessing}.

\subsection{O--H bond dissociation in polymer additives}

\begin{figure}[htp]
    \centering
    \includegraphics[width=0.8\linewidth, bb=0 0 278 91]{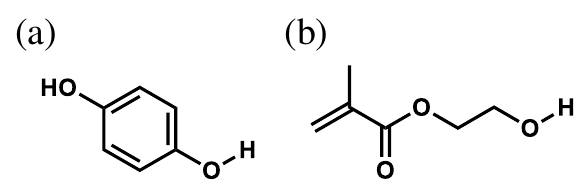}
    \caption{Structural formulas of (a) hydroquinone (HQ) and (b) 2-hydroxyethyl methacrylate (HEMA).}
    \label{fig:scav}
\end{figure}

Finally, we examined the hydrogen dissociation from an O--H bond in three molecules: H$_2$O and two polymer additives (Figure~\ref{fig:scav}).

\begin{figure}[htp]
    \centering
    \includegraphics[width=1.0\linewidth, bb=0 0 1080 2380]{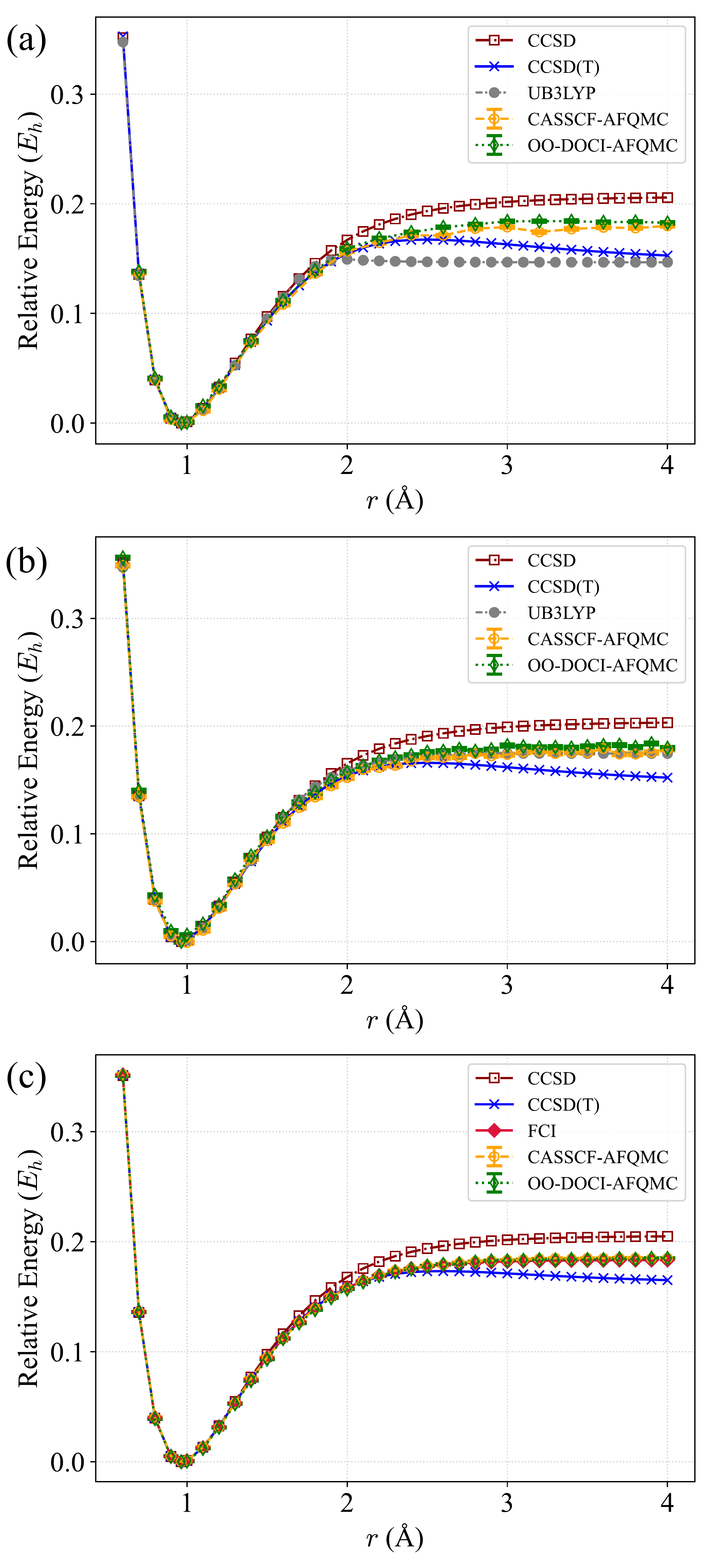}
    \caption{Relative energy curves of (a) HQ, (b) HEMA, and (c) H$_2$O. The 6-31++G$**$ basis set was used for HQ and HEMA, and the cc-pVDZ basis set for H$_2$O. For CASSCF and OO-DOCI, the (8e, 8o) active space was employed for HQ, whereas the (2e, 2o) for HEMA and H$_2$O. CCSD natural orbitals were employed for these MCSCF calculations. The variable $r$ on the horizontal axis denotes the O--H bond distance.}
    \label{fig:OH_CCSD}
\end{figure}

The O--H dissociation energy curves for these molecules are shown in Figure~\ref{fig:OH_CCSD}.
The relative energies were plotted \textcolor{black}{using the minimum energy point of unrestricted B3LYP (UB3LYP)} serving as the reference.
The molecular geometries used in the Gaussian and PySCF calculations differed only by an amount on the order of round-off error and were therefore considered identical.
Standard errors for the ph-AFQMC results are shown as error bars, encompassing the propagated error from the respective reference point.

Among these three molecules examined, the bond dissociation energy $\Delta E$, \textcolor{black}{which is defined as the energy difference between the longest bond length point and the reference point,} followed the trend CCSD(T) $<$ ph-AFQMC $<$ CCSD.
For HQ and HEMA, OO-DOCI-AFQMC aligned closely with CASSCF-AFQMC.
For H$_2$O, OO-DOCI-AFQMC closely mirrored CASSCF-AFQMC and FCI results across the entire range of Figure~\ref{fig:OH_CCSD}~(c). 
Hence, OO-DOCI-AFQMC provides a highly accurate description of O--H bond dissociation.

Conversely, CCSD overestimated the bond dissociation energy, whereas CCSD(T) underestimated it.
For all three molecules, the CCSD(T) energy gradually decreased in the large $r$ region, indicating unphysical behavior. This issue with CCSD(T) will be further discussed in this study.

Additionally, UB3LYP calculations were performed for HQ and HEMA, revealing differing trends for these two molecules.
It closely aligned with ph-AFQMC for HEMA but significantly underestimated the dissociation energy for HQ.

\begin{figure}[htp]
    \centering
    \includegraphics[width=1.0\linewidth, bb= 0 0 2170 2430]{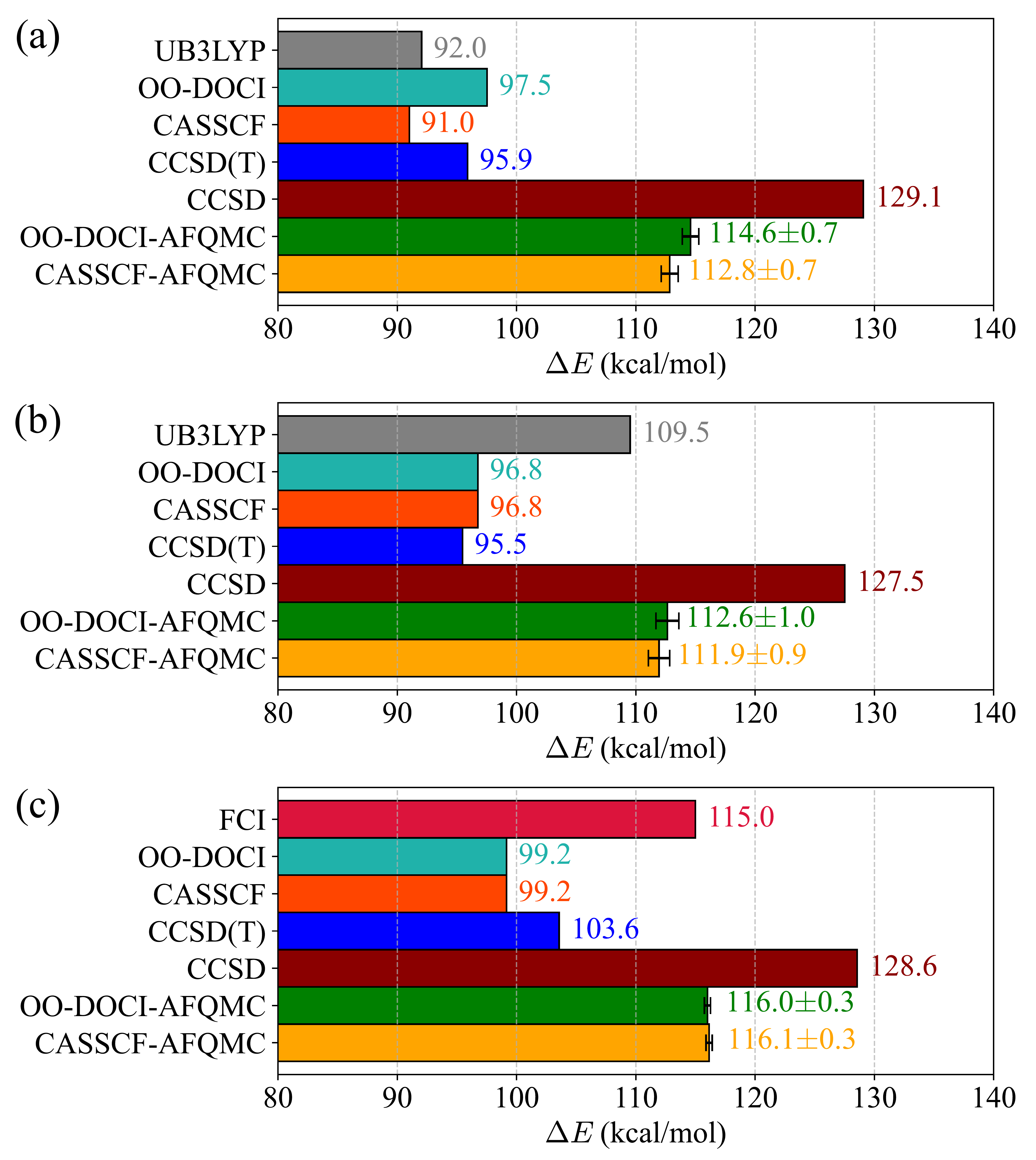}
    \caption{O--H bond dissociation energies for (a) HQ, (b) HEMA, and (c) H$_2$O.}
    \label{fig:OH}
\end{figure}

A summary of the bond dissociation energies $\Delta E$ for each molecule is shown in Figure~\ref{fig:OH}.
The OO-DOCI-AFQMC and CASSCF-AFQMC methods demonstrated excellent agreement and reproduced the FCI result for H$_2$O.
These \textcolor{black}{results} highlight the predictive accuracy of OO-DOCI-AFQMC in determining \textcolor{black}{the} O--H bond dissociation energy, \textcolor{black}{which is} typically around 112~kcal/mol~\cite{2023Bond}.

CASSCF and OO-DOCI underestimated $\Delta E$ by more than 10~kcal/mol, underscoring the importance of recovering \textcolor{black}{dynamical} electron correlation through ph-AFQMC.
For HEMA and H$_2$O, OO-DOCI closely mirrored CASSCF due to their small (2e, 2o) active space.
For HQ, however, OO-DOCI overestimated the CASSCF energy by approximately 6.5~kcal/mol due to the use of larger active space (8e, 8o), including not only the $\sigma$/$\sigma^*$ orbitals of the OH bond but also the $\pi$/$\pi^*$ orbitals of the nearby aromatic ring.

To gain deeper insight into the anomalous behavior of CCSD(T) in the dissociation region, we examine the $T_1$ diagnostic, which measures the degree of static correlation.
A $T_1$ value exceeding $0.02$ indicates that multi-reference characters can not be ignored~\cite{Lee1989diagnostic}.

\begin{figure}[ht]
    \centering
    \includegraphics[width=0.9\linewidth, bb=0 0 492 355]{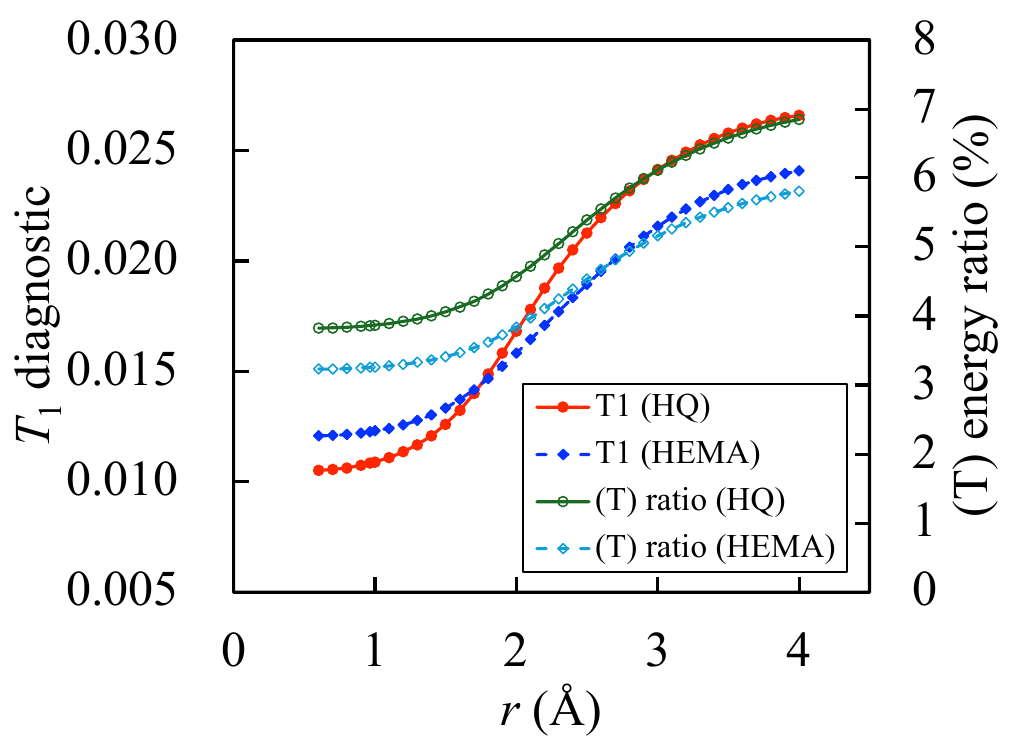}
    \caption{$T_1$ diagnostic and (T) correlation energy ratios for HQ and HEMA.}
    \label{fig:t1}
\end{figure}

The $T_1$ values for HQ and HEMA are shown in Figure~\ref{fig:t1}. 
In both molecules, $T_1$ increased with the bond length $r$ and eventually surpassed 0.02.
Similar trends are observed with the 6-311++G$**$ basis set, although the results are omitted.

The (T) term is a perturbative correction that accounts for triple excitation contributions.
As shown in Figure~\ref{fig:t1}, the fraction of the total correlation energy stemming from (T) also increased with $r$, reflecting the growing reliance on this perturbative correction.
Combined with the increasing $T_1$, these observations underline the limitations of an SR approach in CCSD(T) for describing bond dissociation, thus highlighting the need for MR methods.

\begin{figure}[ht]
    \centering
    \includegraphics[width=0.9\linewidth, bb=0 0 425 326]{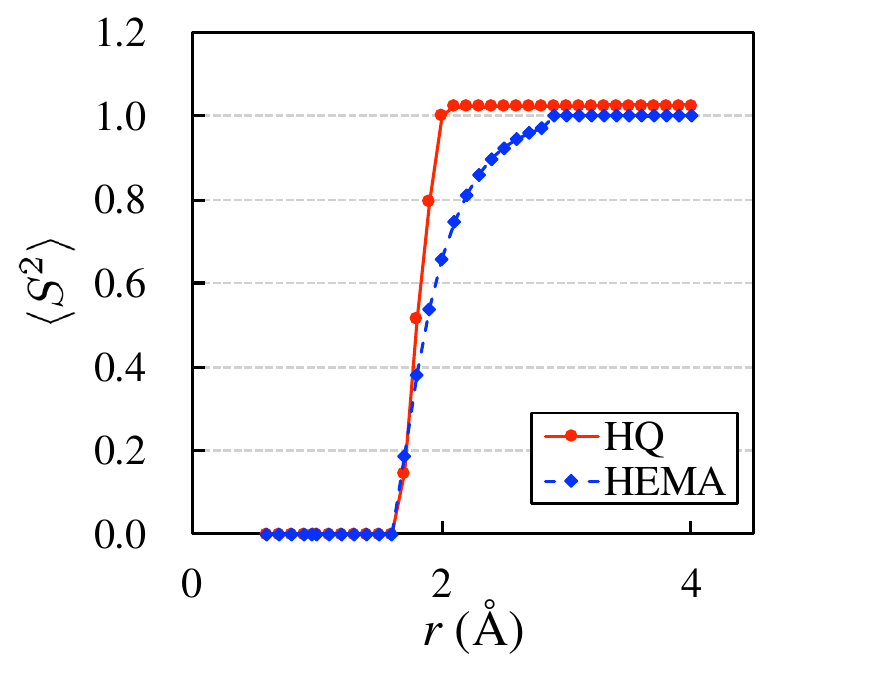}
    \caption{Spin contamination in UB3LYP for HQ and HEMA, measured by $\langle S^2 \rangle$.}
    \label{fig:s2}
\end{figure}

Finally, while the unrestricted approach in UB3LYP can qualitatively treat single-bond dissociation by allowing spin symmetry breaking, the resulting wave function is not necessarily an eigenfunction of the spin-squared operator $S^2$.
Figure~\ref{fig:s2} plots $\langle S^2 \rangle $ for UB3LYP wave functions of HQ and HEMA, revealing the spin contamination at large $r$.
In summary, although UB3LYP can handle single-bond dissociation to a certain extent, it is plagued by spin contamination issues. Conversely, our MR methods provide a more reliable description of O--H bond dissociation.

\section{Conclusions} \label{sec:conclusions}

This study proposed DOCI-/OO-DOCI-AFQMC methods that utilized DOCI/OO-DOCI wave functions as the trial wave function for ph-AFQMC.
These seniority-zero wave functions efficiently captured static electron correlation using only the ground-state and the pair-excited configurations, thereby requiring significantly fewer electronic configurations than CAS-based methods. 

This approach allowed ph-AFQMC to use MR trial wave functions beyond the typical limit of about \textcolor{black}{14} electrons in \textcolor{black}{14} orbitals by reducing the exponential overhead associated with CAS methods~\textcolor{black}{\cite{Couty1997generalized}, as demonstrated by OO-DOCI(12e, 28o)-AFQMC for the carbon dimer}.
A key objective was to investigate the extent to which ph-AFQMC can recover the missing correlation when the trial wave function is restricted to the seniority-zero space.

Numerical results demonstrated that OO-DOCI-AFQMC closely aligned with CASSCF-AFQMC for single O--H bond dissociation and showed good agreement with FCI for the H$_2$O scenario, suggesting reliability for other single-bond dissociation processes.

For multiple bond dissociations, DOCI-/OO-DOCI-AFQMC successfully reproduced CASCI-/CASSCF-AFQMC near equilibrium and at intermediate bond lengths, except for the carbon dimer case. However, ph-AFQMC calculations using \textcolor{black}{these} seniority-zero trial wave functions may lack quantitative accuracy in the fully dissociated regime, highlighting limitations for strongly correlated systems. 

\textcolor{black}{While seniority-zero wave functions are well-known for capturing strong electron correlation, how best to include finite-space configuration interactions that provide an appropriate bias for ph-AFQMC is still an open question.}
Enlarging the seniority space can be a crucial way to improve the accuracy of ph-AFQMC based on MR wave functions, particularly for \textcolor{black}{strongly correlated systems such as} multiple-bond dissociation. 

\textcolor{black}{In terms of further scaling reduction, the use of} other seniority-zero methods~\cite{Limacher2013new, Boguslawski2014efficient, Stein2014seniority,Henderson2014seniority,Henderson2015pair,Tecmer2022geminal,Kawasaki2025low}, \textcolor{black}{is considered as a valuable research direction~\cite{Leszczyk2022assessing} in line of this study}.
\textcolor{black}{In particular, it has been numerically shown that the pair coupled-cluster doubles (pCCD) yields the correlation energy almost identical to that of DOCI, even for system sizes that conventional DOCI cannot reach~\cite{Shepherd2016using}.}
\textcolor{black}{Given the critical importance of the approach that balances electron-paired static and dynamical correlations~\cite{Kobayashi2010generalized,Limacher2014simple,Boguslawski2015linearlized,Boguslawski2017benchmark,Leszczyk2022assessing,Henderson2014seniority,Garza2015synergy,Garza2015range,Garza2015actinide}, the application of the DOCI-AFQMC methodology} to other MR systems requiring larger active spaces, such as polynuclear transition metal complexes, holds promise for future research.

\section*{Acknowledgements}

We thank Nobuki Inoue for fruitful discussion.
We also thank Shigeki Furukawa for providing experimental data on polymer additives. 
This project was supported by funding from the MEXT Quantum Leap Flagship Program (MEXTQLEAP) through Grant No. JPMXS0120319794, and the JST COI-NEXT Program through Grant No. JPMJPF2014.
The completion of this research was partially facilitated by the JSPS Grants-in-Aid for Scientific Research (KAKENHI), specifically Grant Nos. JP23H03819.
Some of the results were obtained using the computational resources of the AI Bridging Cloud Infrastructure (ABCI) provided by National Institute of Advanced Industrial Science and Technology (AIST).


\bibliography{apssamp}

\end{document}